\documentclass[aps,prL,twocolumn,superscriptaddress]{revtex4-1}
\usepackage{graphicx}
\usepackage{dcolumn}
\usepackage{bm}
\usepackage{amsmath}
\usepackage[english]{babel}
\usepackage[utf8]{inputenc}
\usepackage{verbatim}

\usepackage{xcolor}

\usepackage{layouts}

\begin{document}
	\bibliographystyle{achesmo.bst}
	\title{Spin relaxation benchmarks and individual qubit addressability for holes in quantum dots}

	\author{W. I. L. Lawrie}
	\email{w.i.l.lawrie@tudelft.nl}
	\affiliation{QuTech and Kavli Institute of Nanoscience, Delft University of Technology, Lorentzweg 1, 2628 CJ Delft, The Netherlands}
	\author{N. W. Hendrickx}
	\affiliation{QuTech and Kavli Institute of Nanoscience, Delft University of Technology, Lorentzweg 1, 2628 CJ Delft, The Netherlands}
	\author{F. van Riggelen}
	\affiliation{QuTech and Kavli Institute of Nanoscience, Delft University of Technology, Lorentzweg 1, 2628 CJ Delft, The Netherlands}
    \author{M.~Russ}
	\affiliation{QuTech and Kavli Institute of Nanoscience, Delft University of Technology, Lorentzweg 1, 2628 CJ Delft, The Netherlands}
	\author{L. Petit}
	\affiliation{QuTech and Kavli Institute of Nanoscience, Delft University of Technology, Lorentzweg 1, 2628 CJ Delft, The Netherlands}
	\author{A. Sammak}
	\affiliation{QuTech and Netherlands Organization for Applied Scientific Research (TNO), Stieltjesweg 1 2628 CK Delft, The Netherlands}
	\author{G. Scappucci}
	\affiliation{QuTech and Kavli Institute of Nanoscience, Delft University of Technology, Lorentzweg 1, 2628 CJ Delft, The Netherlands}
	\author{M. Veldhorst}
	\email{m.veldhorst@tudelft.nl}
	\affiliation{QuTech and Kavli Institute of Nanoscience, Delft University of Technology, Lorentzweg 1, 2628 CJ Delft, The Netherlands}

	\begin{abstract}
    We investigate hole spin relaxation in the single- and multi-hole regime in a 2x2 germanium quantum dot array. We use radiofrequency (rf) charge sensing and observe Pauli Spin-Blockade (PSB) for every second interdot transition up to the (1,5)-(0,6) anticrossing, consistent with a standard Fock-Darwin spectrum. We find spin relaxation times $T_1$ as high as 32 ms for a quantum dot with single-hole occupation and 1.2 ms for a quantum dot occupied by five-holes, setting benchmarks for spin relaxation times for hole quantum dots. Furthermore, we investigate the qubit addressability and sensitivity to electric fields by measuring the resonance frequency dependence of each qubit on gate voltages. We are able to tune the resonance frequency over a large range for both the single and multi-hole qubit. Simultaneously, we find that the resonance frequencies are only weakly dependent on neighbouring gates, and in particular the five-hole qubit resonance frequency is more than twenty times as sensitive to its corresponding plunger gate. The excellent individual qubit tunability and long spin relaxation times make holes in germanium promising for addressable and high-fidelity spin qubits in dense two-dimensional quantum dot arrays for large-scale quantum information.
	\end{abstract}
	\maketitle
	
	\begin{figure}[h!]
	\includegraphics[width = \linewidth]{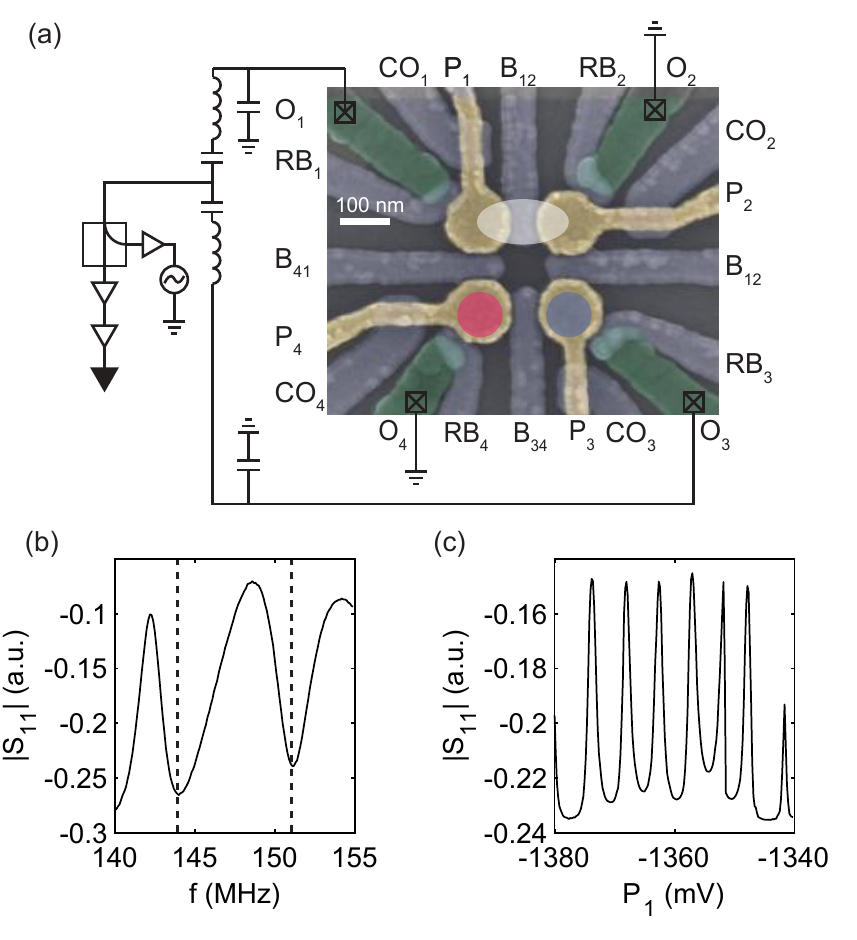}%
	\caption{(a) Coloured scanning electron microscope image of a nominally identical 2x2 quantum dot array. Each quantum dot is defined by a plunger gate, P$_{1-4}$ (yellow) and barrier gates B$_{12-41}$ (blue) are used to set the tunnel coupling. In addition, each quantum dot is coupled to a reservoir, O$_{1-4}$ (green), via a barrier gate RB$_{1-4}$. A cut off gate, CO$_{1-4}$, is present for good confinement of the quantum dots. Ohmics 1 and 3 are bonded to an inductor to create a tank circuit with the parasitic capacitance of the device to ground. A radiofrequency tone is applied to the ohmics and the reflected signal $|\Delta S_{11}|$ returns via a directional coupler and is read out. (b) Reflected signal of the two tank circuits. Two clear resonances occur at \textit{f}$_{\text{O}1}$ = 150.7 MHz and \textit{f}$_{\text{O}3}$ = 143.3 MHz for tank circuits connected to ohmics O$_1$ and O$_3$ respectively. (c) Single-hole transistor (SHT) Coulomb oscillations measured in the tank circuit response by applying a microwave tone of 150.7 MHz. A sensing quantum dot is formed underneath the plunger gates P$_1$ and P$_2$, by opening the interdot barrier gate B$_{12}$.}
    \end{figure}
    
	Qubits based on spin states are well established candidates for quantum information processing \cite{Loss1998}. Pioneering studies were conducted on low-disorder gallium arsenide heterostructures \cite{Petta2005, Koppens2006}, but quantum coherence remained limited due to hyperfine interaction with nuclear spins. These interactions can be eliminated by using isotopically enriched group IV semiconductors as the host material \cite{Itoh2014}. In silicon this has led to landmark achievements, such as extremely long quantum coherence \cite{Veldhorst2014} and relaxation times \cite{Yang2013}, single qubit gates with fidelities beyond 99.9$\%$ \cite{Yang2019a,Yoneda2018}, execution of two-qubit gates \cite{Veldhorst2015N, Zajac2018}, quantum algorithms \cite{Watson2018}, and the operation of single qubit rotations \cite{Yang2019} and two-qubit logic \cite{Petit2019a} above one Kelvin as a key step toward quantum integrated circuits \cite{Vandersypen2017, Veldhorst2017, Li2018}.

    In it's natural form, germanium contains only 7.76\% isotopes with non-zero nuclear spin and, like silicon, can be isotopically enriched \cite{Itoh1993} to eliminate nuclear spin dephasing. Recent advances in materials science enabled high mobility strained planar germanium (Ge/SiGe) heterostructures \cite{Sammak2019} for the fabrication of stable gate-defined quantum dots that can confine holes \cite{Hendrickx2018}, which are predicted to have a multitude of favourable properties for quantum control \cite{Bulaev2005, Bulaev2007}. The inherent strong spin-orbit coupling of holes allows for fast qubit control \cite{Maurand2016, Watzinger2018, Hendrickx2020} without integrating external components that complicate scalability, such as nano-magnets and microwave antennas. Moreover, holes do not suffer from valley degeneracy and their small effective mass of m$^{*}_{\text{h}}$ = 0.05 m$^{*}_{\text{e}}$ \cite{Lodari2019} gives rise to large orbital splittings at the band center. These beneficial aspects thereby position holes in germanium as a promising material for quantum information \cite{Scappucci2020}.
	
\begin{figure*}
	\includegraphics[]{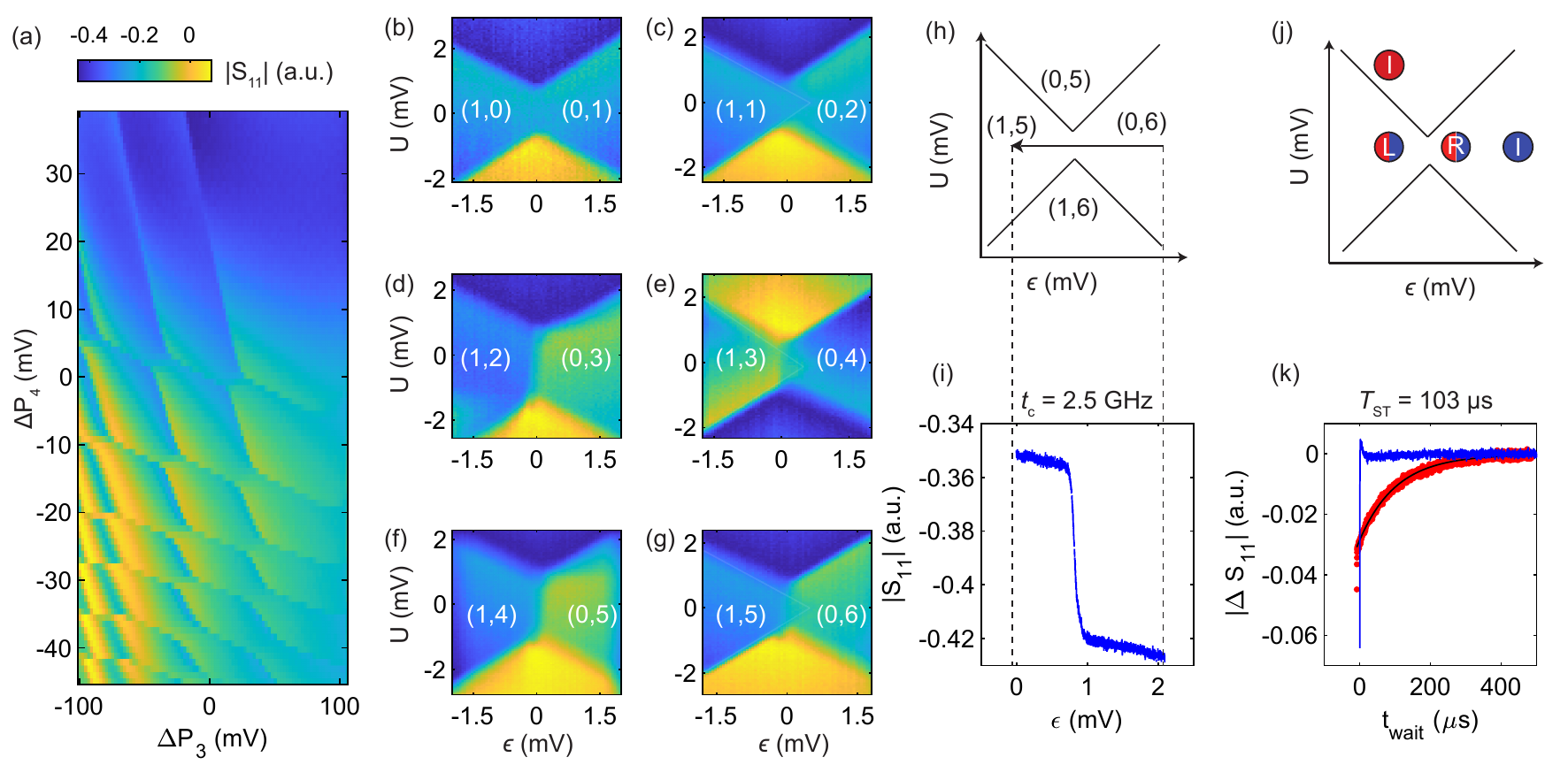}%
	\caption{(a) Double quantum dot charge stability diagram obtained by rf-charge sensing in a reconfigurable quadruple quantum dot. A double quantum dot is formed under P$_3$ (blue) and P$_4$ (red) with controllable interdot tunnel coupling by tuning interdot barrier gate B$_{34}$ and using rf-charge sensing we can clearly monitor the charge occupancy. The map is taken with 2000 averages. (b-g) Charge stability diagrams of the charge transitions highlighted in (a) showing spin blockade between (1,N-1)$\Longleftrightarrow$(0,N) transitions only when N is even (2,4,6) following expected Fock-Darwin filling. White lines mark the expected spin blockade regions in c,e,g. (h-i) Interdot tunnel coupling measurement for the (1,5)$\Longleftrightarrow$(0,6) transition. Sweep direction is negative in detuning axis to avoid spin blockade artifacts in spectrum. A tunnel coupling of \textit{t}$_c$ = 2.5 GHz is obtained and kept within 200 MHz of this value for all measurements in the work. (j-k) Singlet-triplet readout traces, showing signal difference between the spin-blocked and unblocked states. Each trace is averaged 1000 times. Loading a random spin under P$_3$ (red) leads to a singlet-triplet decay of \textit{T}$_{\text{ST}}$ = 103 $\mu$s. }
\end{figure*}
		
	    \begin{figure}
	\includegraphics[width = \linewidth]{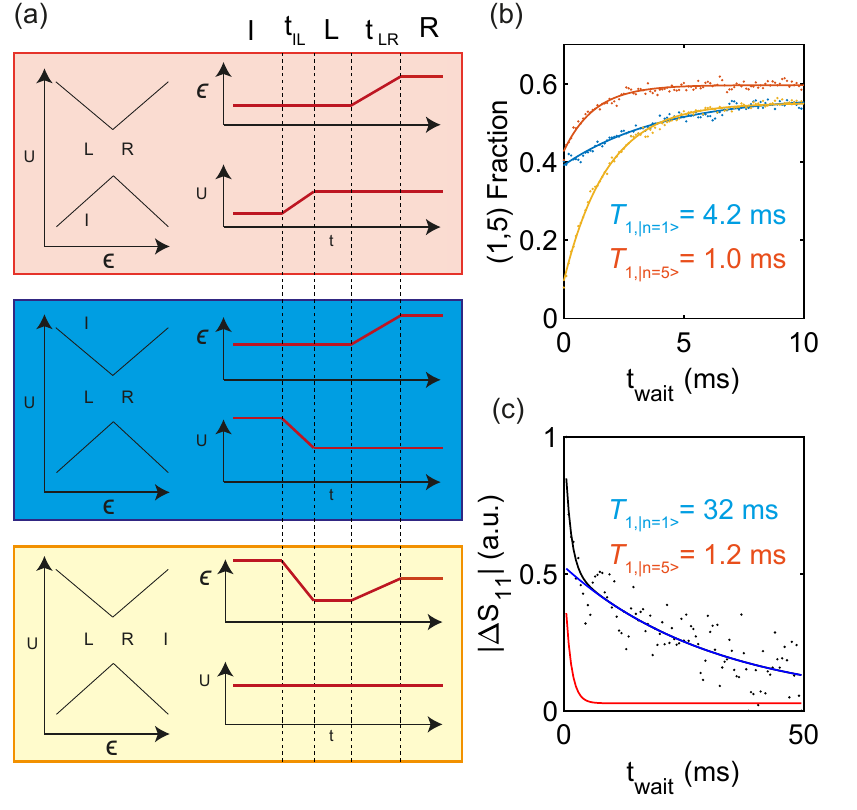}%
	\caption{(a) Pulse sequences utilized for different loading protocols. Red loads a random spin in P$_4$, blue loads a random spin in P$_3$, and yellow loads either the $|\text{S}_{(1,5)}\rangle$ or a mixture of $|\text{S}_{(1,5)}\rangle$ and $|\text{T}_{0}\rangle$ depending on the adiabaticity of the IL pulse.  (b) Deterministic loading of a single-hole in P$_4$ and P$_3$ (red, blue respectively) and mixed state loading (yellow). We extract spin relaxation times of T$_{1,|n=5>}$ =  1.0 ms and T$_{1,|n=1>}$ 4.23 ms for each deterministically loaded quantum dot, which we fit as a double exponential for the mixed loading case. (c) Longest spin relaxation trace taken after minimizing reservoir-dot tunnel coupling. We extract two spin relaxation times of T$_{1,|n=1>}$ = 32 ms and T$_{1,|n=5>}$ = 1.2 ms. }
    \end{figure}
            
    \begin{figure}[t]
	\includegraphics[width = \linewidth]{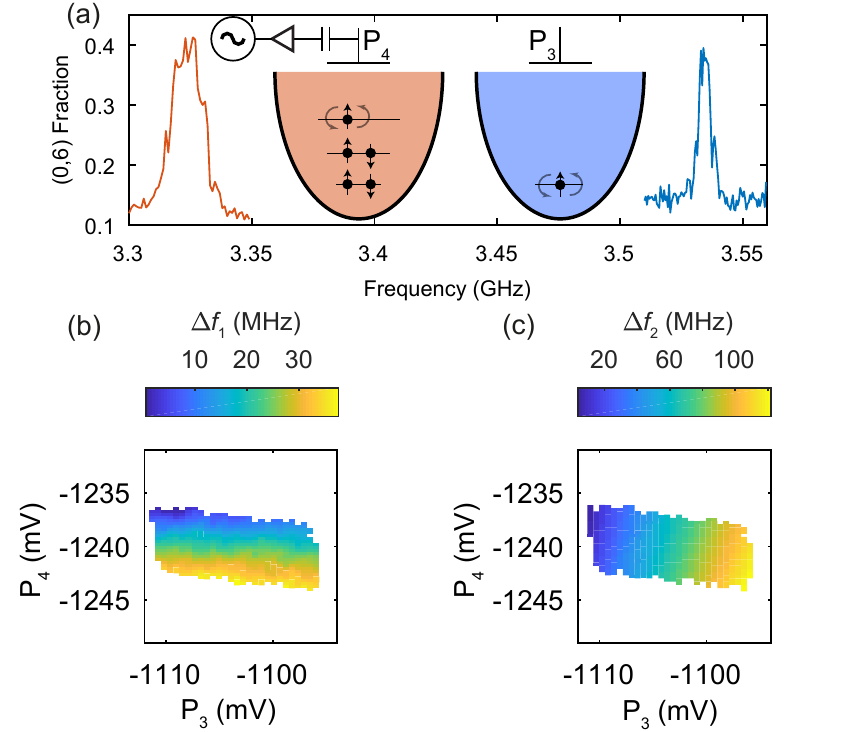}%
	\caption{(a) Qubit resonance frequency of the five-hole (3.33 GHz) and single-hole qubit (3.53 GHz). The magnetic field is set to B = 667 mT. We extract in-plane $g-$factors of $g_{|n=1\rangle}$ = 0.362 and $g_{|n=5\rangle}$ = 0.383 for plunger gate values P$_3$ = 1098 mV,  P$_4$ = 1236 mV. (b) Five-hole (\textit{f}$_1$) and (c) single-hole (\textit{f}$_2$) resonance frequency dependence on gate voltage. We find a strong dependence of the resonance frequency on the respective plunger gate, but a significantly reduced dependence on the neighbouring plunger gate voltage.}
    \end{figure}
	While it has been demonstrated that both single- and multi-hole qubits can be coherently controlled and read out in planar germanium \cite{Hendrickx2019b,Hendrickx2020}, an open question remains which hole occupancy is most advantageous for quantum operation. Electron spin qubits in silicon have been operated with quantum dots containing one, three and even more electrons, with more electrons typically performing favourably in terms of driving speed when driven electrically due to greater wave function mobility \cite{Veldhorst2015P,Leon2020}. Here, we focus on single and multi-hole spin qubit operation in germanium and concentrate on two critical elements for quantum information with quantum dots: the spin relaxation time and the qubit addressability. We find that both the spin relaxation times of the single-hole ($T_{1,|n=1\rangle}$) and five-hole ($T_{1,|n=5\rangle}$) qubits are long, with the longest relaxation time for single-holes measured to be $T_{1,|n=1\rangle} = 32$ ms. Furthermore, we observe that single and multi-hole qubits exhibit a strong but comparable resonance frequency dependence on electric gate voltage. Interestingly, we find that while the qubit resonance frequency can be significantly tuned with the corresponding plunger gate, it is only weakly dependent on neighbour plunger gates. We thereby conclude that hole spin qubits can be locally addressed, crucial for the operation of dense qubit arrays.\\
	\newline

    The experiments are performed on a two-dimensional 2x2 quantum dot array fabricated using a multi-layer gate stack \cite{Lawrie2020} (See Fig 1a). Four plunger gates P$_{1-4}$ define four quantum dots, whose interdot tunnel couplings are controllable via barrier gates B$_{12-41}$. Four metallic reservoirs O$_{1-4}$ can be controllably coupled to each quantum dot via their respective barrier gates RB$_{1-4}$. We operate in a configuration whereby electrostatic gates P$_1$, B$_{12}$ and P$_2$ define one large quantum dot, serving as single-hole transistor (SHT) for charge sensing, shown in Fig 1b. By connecting an inline NbTiN kinetic inductor of L $\approx$ 2 $\mu$H to the ohmic O$_1$, we form a resonant tank circuit at a frequency of \textit{f}$_{\text{O}1}$ =  150.7 MHz used for fast rf charge sensing \cite{Schoelkopf1998,Vukusic2017a}. A second inline inductance connected to ohmic O$_3$ makes the device reconfigurable for any of the four double quantum dot-sensor combinations, with a second tank circuit resonance at \textit{f}$_{\text{O}3}$ = 143.3 MHz. The reflected signal response to rf power delivered to the sample in a frequency range encompassing these two resonances is shown in fig 1b. Modulation of the channel resistance due to Coulomb oscillations in the SHT is shown in Fig 1d. Next, we apply voltages to the plunger gates P$_3$ and P$_4$ to form quantum dots that load via the reservoir barriers RB$_3$ and RB$_4$, respectively. By applying sawtooth wave pulses to the plunger gates and simultaneously applying the inverse pulses to the SHT plunger gates P$_1$ and P$_2$, we can tune up the device to a double quantum dot of arbitrary occupancy while compensating the charge sensor in real time.\\
    
    Figure 2a shows the charge stability diagram for a sweep of gates P$_3$ versus P$_4$ over the first few charge addition lines in each quantum dot.
	We focus on the set of anticrossings for the first charge addition line of the quantum dot under P$_3$ of the form (N$_{P_3}$, \text{N}$_{P_4}$)\, $\Longleftrightarrow$ (N$_{P_3}$-1, \text{N}$_{P_4}+1)$, where N$_{P_{3(4)}}$ is the charge occupation of the quantum dot formed under P$_{3(4)}$, from double dot occupation (1,0)\, $\Longleftrightarrow$ (0,1) to (1,5) $\Longleftrightarrow$ (0,6).
	We define a virtual gate space in detuning $\epsilon$ and energy U through a linear transformation of the gate voltages on P$_3$ and P$_4$. We apply sawtooth wave pulses that sweep $\epsilon$ from -2 mV to +2 mV and steps U from -2.5 mV to +2.5 mV with respect to the anticrossing. Figures 2b-g shows the resulting stability diagrams.

     Pauli spin blockade is observed for (N$_{\text{odd}}$,N$_{\text{odd}}$) $\Longleftrightarrow$ (N$_{\text{even}}$,N$_{\text{even}}$) type transitions up to the sixth occupancy. This is consistent with a Fock-Darwin level filling observed for electrons in gallium arsenide \cite{Tarucha1996} and holes in silicon \cite{Liles2018a} until the same charge occupancy. Working now with the (1,5) $\Longleftrightarrow$ (0,6) anticrossing, we extract a lever arm $\alpha=$ 0.18 from the thermally broadened polarization line using a hole temperature of 100 mK, which allows us to extract the tunnel couplings \cite{DiCarlo2004}. Figures 2h,i show the used pulse scheme and measured trace. The interdot coupling is kept constant within \textit{t}$_c$ = 2.5 $\pm$ 0.2 GHz for all measurements. Figures 2j,k show the relevant pulse sequences and resulting traces for Pauli spin blockade (PSB) readout. Here we compare the partially blocked (red) and unblocked (dark blue) signals, allowing us to distinguish between the spin up and spin down states in the lower energy quantum dot.
     By loading a hole in P$_3$ with a random spin state, we expect to observe a blocked signal approximately half of the time. Monitoring the readout signal in the charge sensor as a function of time provides us with the spin relaxation at the readout position, which we find to be \textit{T}$_{\text{ST}}$ = 103 $\mu$s. \\
     \newline
    
	We now assess the spin relaxation of the single- and five-hole qubits. All experiments were performed at a magnetic field $B$ = 0.67 T, allowing for a comparison with previous germanium hole spin qubit experiments \cite{Hendrickx2020, Hendrickx2019b} in a similar magnetic field regime. Figure 3a shows the pulse sequences used to measure the spin relaxation times of the hole spins in each quantum dot. Each pulse sequence consists of an initialization (I), load (L), and read (R) phase, with two ramps between the I and L phases (t$_{\text{IL}}$), and the L and R phases (t$_\text{LR}$). Using the first two sequences (red and blue in Fig. 3a) a randomly orientated spin is loaded into the quantum dot defined under P$_4$ or P$_3$ respectively. This allows deterministic probing of the spin relaxation time of each dot, by varying the load wait time t$_L$.

	The third pulse sequence (yellow in Fig. 3a) initializes the system in the singlet state with charge configuration ($N_{P3}$,$N_{P4}$) = (0,6) ($|\text{S}_{(0,6)}\rangle$). The system is then tuned to the charge configuration ($N_{P3}$,$N_{P4}$) = (1,5)  ($|\text{S}_{(1,5)}\rangle$). We pulse with a ramp time t$_{IL}$ = 100 ns, resulting in a diabatic movement through the charge section, and through fast charge relaxation we expect to initialize the $|\uparrow,\downarrow\rangle$ and $|\downarrow,\uparrow\rangle$ states randomly with equal probability. This initialization then allows us to efficiently measure both spin relaxation times in a single measurement and is useful since it allows for fast measurements even when the quantum dot-reservoir couplings are low.
	
	In Fig. 3b, we show the spin relaxation times of the quantum dots using the three sequences. We find $T_{1,|n=5\rangle}$ = 1.0 ms and $T_{1,|n=1\rangle}$ = 4.23 ms by fitting exponential decays to the individual measurements. The measurement corresponding to the sequence with randomly preparing a spin up state in one of the two quantum dots is fitted with a double exponential curve using the time constants of the individual decays, and we have left the amplitudes and asymptotes as free fitting parameters. We find approximately equal amplitudes for each decay, in correspondence with an equal loading of both anti-parallel spin states.
	
	We can further increase the single-hole relaxation time by reducing the quantum dot-reservoir coupling. Using the barrier gate RB$_3$, we tune the quantum dot-reservoir coupling of the single-hole quantum dot from 81.43 KHz to 27.45 KHz (see supporting information section I). The spin relaxation decay shown in Fig. 3c has been analysed using the above mentioned double exponential fit and we find an significantly increased single-hole spin relaxation time $T_{1,|n=1\rangle}$ = 32 ms. This spin relaxation time is significantly longer than results reported for planar germanium quantum dots (\textit{T}$_{1,|n=1\rangle}$ = 1.2 ms \cite{Hendrickx2019b}), hut wires (\textit{T}$_1$ = 90 $\mu$s \cite{Vukusic2018}), nanowires (\textit{T}$_1$ = 600 $\mu$s \cite{Hu2012}), and even holes in gallium arsenide (\textit{T}$_1$ = 60 $\mu$s \cite{Bogan2019}) and silicon (\textit{T}$_1$ = 8.3 $\mu$s \cite{Bohuslavskyi2016}) at similar magnetic fields. Spin states in planar germanium thereby define the benchmark for spin relaxation in hole based quantum dots.\\
 \newline
 
The presence of spin-orbit coupling allows for electrical and coherent control of the spin states without the need for additional structures such as striplines or micromagnets \cite{Bulaev2005, Bulaev2007, Hendrickx2020,Watzinger2018}. We investigate the individual tunability and addressability of the single- and multi-hole qubits. In Fig. 4a, we show results where we have applied a microwave tone of length $t_{\text{mw}}$ = 400 ns to the gate P$_4$. We observe two resonance frequencies at 3.33 GHz and 3.53 GHz in Fig 4a, corresponding to an in-plane Zeeman energy difference d\textit{E}$_{z}$ = 200 MHz. Figures 4c,d show the dependence of each resonance frequency on the electrostatic gate voltages on the two relevant plunger gates P$_3$ and P$_4$. We initialize in the $|\text{S}_{(0,6)}\rangle$ singlet state, then load in different points in the (1,5) charge state by changing the potentials applied to P$_3$ and P$_4$. We then manipulate the spins by applying a microwave tone to P$_4$ and read out in the PSB window. The resonance frequency dependence on gate voltage is approximately linear. For the five-hole qubit we find a dependence on its plunger gate voltage d\textit{f}$_1$/dP$_4$ = -4.78 MHz/mV and we find d\textit{f}$_1$/dP$_3$ = -0.155 MHz/mV. For the single-hole qubit we find a slightly stronger dependence on its plunger gate voltage d\textit{f}$_2$/dP$_3$ = 6.78 MHz/mV and we find a cross talk, d\textit{f}$_2$/dP$_4$ = -1.79 MHz/mV.
This corresponds to a cross talk ratio of about 1/30 for the five-hole qubit and about 1/4 for the single-hole qubit. The cross talk for the single-hole qubit is comparable to the lever arm ratio (see supporting information section II) $\alpha_{P_3/P_4}(f_2)$ = 0.11. Remarkably, the five-hole qubit has a lever arm ratio $\alpha_{P_4/P_3}(f_1)$ = 0.07, significantly larger then the resonance frequency cross talk ratio.\\
\newline

 In summary, we have demonstrated benchmarks for spin relaxation in hole quantum dots and found $T_{1,|n=1\rangle}$ = 32 ms for a single-hole qubit and $T_{1,|n=5\rangle}$ = 1.2 ms for a five-hole qubit and conclude that spin relaxation is not a bottleneck for quantum computation with holes. We have shown the presence of Pauli-spin blockade at different hole fillings and found it to be consistent with a Fock-Darwin spectrum that only involves spin degeneracy. We find that both the single-hole and multi-hole qubit resonance frequency can be tuned over a large range. We find that the resonance frequencies are only weakly dependent on neighbouring gates, which results in good local addressability. The observation of the sign difference in the resonance frequency dependence on gate voltage and the strength of the cross talk ratio of the resonance frequencies may provide insights in the nature of the driving mechanism of holes in planar germanium. This is relevant for future work and a possible scenario is that the reduced cross talk of the five-hole qubit originates from an increased heavy-hole light-hole mixing. Such a change may affect the qubit resonance frequency dependence on the amplitude and orientation of the electric field, but further research is needed to investigate this. The long spin lifetimes and excellent individual qubit addressability are encouraging for the operation of hole qubits positioned in large two-dimensional arrays. 
 
\section*{Acknowledgements} 
The authors acknowledge support through a FOM Projectruimte, associated with the Netherlands Organisation for Scientific Research (NWO).

 \providecommand{\latin}[1]{#1}
\makeatletter
\providecommand{\doi}
  {\begingroup\let\do\@makeother\dospecials
  \catcode`\{=1 \catcode`\}=2 \doi@aux}
\providecommand{\doi@aux}[1]{\endgroup\texttt{#1}}
\makeatother
\providecommand*\mcitethebibliography{\thebibliography}
\csname @ifundefined\endcsname{endmcitethebibliography}
  {\let\endmcitethebibliography\endthebibliography}{}

\clearpage
\newpage
\onecolumngrid

\renewcommand{\figurename}{Supporting Figure}
\setcounter{figure}{0}  

\begin{center}
	\textbf{\large Supporting Information}
\end{center}
\renewcommand{\bibnumfmt}[1]{[S#1]}
\renewcommand{\citenumfont}[1]{S#1}

	\bibliographystyle{achesmo.bst}
	\title{Spin relaxation benchmarks and individual qubit addressability for holes in quantum dots:\\ Supporting Information}

	\author{W. I. L. Lawrie}
    \email{w.i.l.lawrie@tudelft.nl}
	\affiliation{QuTech and Kavli Institute of Nanoscience, Delft University of Technology, Lorentzweg 1, 2628 CJ Delft, The Netherlands}
	\author{N. W. Hendrickx}
	\affiliation{QuTech and Kavli Institute of Nanoscience, Delft University of Technology, Lorentzweg 1, 2628 CJ Delft, The Netherlands}
	\author{F. van Riggelen}
	\affiliation{QuTech and Kavli Institute of Nanoscience, Delft University of Technology, Lorentzweg 1, 2628 CJ Delft, The Netherlands}
    \author{M.~Russ}
	\affiliation{QuTech and Kavli Institute of Nanoscience, Delft University of Technology, Lorentzweg 1, 2628 CJ Delft, The Netherlands}	
	\author{L. Petit}
	\affiliation{QuTech and Kavli Institute of Nanoscience, Delft University of Technology, Lorentzweg 1, 2628 CJ Delft, The Netherlands}
	\author{A. Sammak}
	\affiliation{QuTech and Netherlands Organization for Applied Scientific Research (TNO), Stieltjesweg 1 2628 CK Delft, The Netherlands}
	\author{G. Scappucci}
	\affiliation{QuTech and Kavli Institute of Nanoscience, Delft University of Technology, Lorentzweg 1, 2628 CJ Delft, The Netherlands}
	\author{M. Veldhorst}
	\email{m.veldhorst@tudelft.nl}
	\affiliation{QuTech and Kavli Institute of Nanoscience, Delft University of Technology, Lorentzweg 1, 2628 CJ Delft, The Netherlands}

	\maketitle

\section{Tunnel Rate Analysis}
We measure the dot-reservoir tunnel coupling of the quantum dot under plunger gate $P_3$. Before each measurement of spin relaxation in Figure 3 of the main text, we pulse from the (0,5) charge state to the measurement point (1,5) charge state, and measure the sensor response. We observe an exponential decay, the time constant of which determines our dot reservoir tunnel coupling. We pulse the virtual energy gate $U$ from the (0,5) to the (1,5) charge state. Supplementary Figures 3a-b show the resulting sensor response as a function of time in the (1,5) charge state for the spin relaxation measured in Main text figures 3b and 3c respectively. Due to the imperfect charge sensor compensation, we observe a short initial transient in the first few microseconds, followed by the actual charge state transient to which we fit an exponential decay, shown in the inset. We extract dot-reservoir load rates of 81.43 kHz to 27.45 kHz  for the spin relaxation times measured in Main text figures 3b and 3c respectively. \\
\newline

\begin{figure}[h!]
    \centering
    \includegraphics{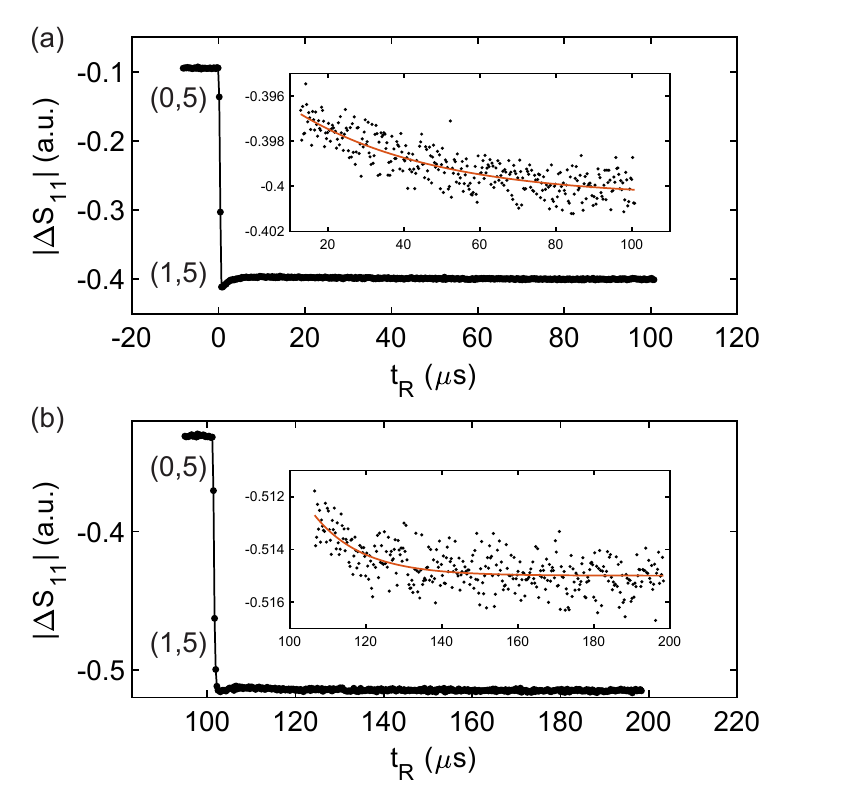}
    \caption{Dot-Reservoir coupling between the singly occupied quantum dot defined under the plunger gate P$_3$. (a) A dot-reservoir coupling of 81.43 kHz is extracted from the dot loading transient for the spin relaxation measurement in in main text figure 3b, and (b) 27.45 kHz for main text figure 3c. }
    \label{fig:my_label}
\end{figure}
\section{Relative Lever Arm}
We extract the relative lever arms of the plunger gates P$_3$ and P$_4$ to the quantum dot potentials underneath them. We take these values from the charge addition line slopes in the stability diagram in figure 2g of the main text. Here, the figure is taken using a virtual gate matrix space of detuning and energy ($\epsilon$,U) which are defined as a linear combination of the voltages on gates P$_3$ and P$_4$ (V$_{P3}$,V$_{P4}$):
\begin{equation*}
\begin{pmatrix}
\epsilon \\
U
\end{pmatrix} =
\begin{pmatrix}
1 & -1.05 \\
1.05 & 1 
\end{pmatrix} 
*
    \begin{pmatrix}
V_{P3} \\
V_{P4} 
\end{pmatrix} 
\end{equation*}
By calculating the gradient of the single and multi hole qubit charge addition lines in main text figure 2g, we can solve for the changes in the plunger gate voltage space, and calculate the ratios for each quantum dot, giving  $\alpha_{P_3/P_4}$ = 0.11 and $\alpha_{P_4/P_3}$ = 0.07.

\end{document}